\documentclass[aastex61]{aastex6}

\begin{document}
\title{Dynamic Spectral Imaging of Decimetric Fiber Bursts \\ in an Eruptive Solar Flare}
\author{Zhitao Wang\altaffilmark{1}, Bin Chen\altaffilmark{1}, Dale E. Gary\altaffilmark{1}}
\affil{\textsuperscript{1} Center for Solar-Terrestrial Research, New Jersey Institute of Technology, University Heights, Newark, NJ 07102}
\email{zw56@njit.edu}

\begin{abstract}

Fiber bursts are a type of fine structure that is often superposed on type IV radio continuum emission during solar flares. Although studied for many decades, its physical exciter, emission mechanism, and association with the flare energy release remain unclear, partly due to the lack of simultaneous imaging observations. We report the first dynamic spectroscopic imaging observations of decimetric fiber bursts, which occurred during the rise phase of a long-duration eruptive flare on 2012 March 3, as obtained by the Karl G. Jansky Very Large Array in 1--2 GHz. Our results show that the fiber sources are located near and above one footpoint of the flare loops. The fiber source and the background continuum source are found to be cospatial and share the same morphology. It is likely that they are associated with nonthermal electrons trapped in the converging magnetic fields near the footpoint, as supported by a persistent coronal hard X-ray source present during the flare rise phase. We analyze three groups of fiber bursts in detail with dynamic imaging spectroscopy and obtain their mean frequency-dependent centroid trajectories in projection. By using a barometric density model and magnetic field based on a potential-field extrapolation, we further reconstruct the 3-D source trajectories of fiber bursts, for comparison with expectations from the whistler wave model and two MHD-based models. We conclude that the observed fiber burst properties are consistent with an exciter moving at the propagation velocity expected for whistler waves, or models that posit similar exciter velocities.

\end{abstract}


\section{Introduction} \label{sec:introduction}

Solar radio bursts in the meter-decimeter wavelengths often reveal various types of narrow frequency-band structure with rapid temporal changes in the dynamic spectra. These spectral fine structures (FS) contain important information on accelerated particles, local plasma and magnetic field properties, and other non-thermal processes in the flaring region \citep{1998ARA&A..36..131B, 2006SSRv..127..195C, 2011ApJ...736...64C}. In particular, fiber bursts have been identified as one of the most prominent types of FS, which often appear as a group of narrowband drifting structures in the dynamic spectrum. Because their frequency drift rates are generally greater than shock-generated type II radio bursts and lower than type III radio bursts produced by fast electron beams, fiber bursts are also known as intermediate drift bursts \citep{1998A&A...333.1034B}, and are believed to be associated with some type of propagating wave excited by flare-accelerated particles.

Observations of fiber bursts have a long history, covering meter and decimeter wavelengths \citep[e.g.,][]{1961ApJ...133..243Y, 1972SoPh...25..210S, 1987SoPh..112..347A, 2005PlPhR..31..314C}. In the decimetric wavelength (dm-$\lambda$) range, the dynamic spectral properties of fiber bursts have been well studied by \citet{1998A&A...333.1034B}, and are briefly reiterated here: (1) dm-$\lambda$ fiber bursts usually have negative frequency drift rates (their peak intensity drifts to lower frequencies in time), suggesting that the emission source propagates toward a lower density region in the solar corona; (2) The relative frequency drift rate $\mid\dot{\nu}/\nu\mid$ is about 0.04--0.1 s$^{-1}$; (3) The time duration at a single frequency is less than 1 s, and the instantaneous frequency bandwidth is less than 2\% of its emission frequency; (4) A parasitic absorption band is usually, but not always, observed below the emission frequency of each burst, although it is weaker than the emitting feature and does not compensate for the emission intensity.

There has been significant progress in the theoretical understanding of fiber bursts in the past several decades.  So far, there are several major types of models concerning the exciter of fiber bursts: whistler wave packets, Alfv\'en solitons, or fast magnetoacoustic sausage-mode disturbances. The whistler wave model \citep{1975SoPh...44..173K} considers the parametric interaction between a propagating whistler wave packet ($w$) and localized Langmuir waves ($L$) excited by electrons with an anisotropic distribution ($L+w\rightarrow t$, where $t$ denotes the electromagnetic radio waves). As a result, some energy is taken from the background continuum and upconverted by an amount equal to the whistler frequency $\omega_w$ to produce the fiber burst emission ridge. This model is successful in interpreting the frequency drift of fiber bursts and the absorption-emission pairs in the spectrum, but it has been criticized for the inefficiency of the emission process under typical coronal conditions \citep[e.g.,][]{1975AuJPh..28..101M}.

To account for the observed flux of fiber bursts, \citet{1983ApJ...264..677B} suggested that a large number of whistler solitons can be produced through some nonlinear processes, although $10^{11}$--$10^{14}$ whistler solitons are needed to account for the observed flux of a single fiber. For these solitons to emit radio waves collectively while propagating, they have to be confined within a coherent density structure, one possible candidate for which is an Alfv\'en wave soliton. \citet{1990A&A...236..242T} focused on the modulation of emission by the Alfv\'en solitons, while leaving the emission process itself open. They suggested that these solitons have super-Alfv\'enic speeds with the Alfv\'enic Mach number between 1 and 3. However, \citet{1987SoPh..110..381M} argued that these whistler solitons would suffer significant cyclotron damping for $x=\omega_\mathrm{w}/\omega_\mathrm{ce}>0.25$ in the corona with a low plasma beta. In addition, the modulation of the background density and emission by the Alfv\'en soliton should lead to the redistribution of emission with respect to the local plasma frequency. Therefore the total emission-absorption intensity should be conserved, which is often inconsistent with the observations.

Recently, \citet{2006SoPh..237..153K} proposed another type of model that involves magnetohydrodynamic (MHD) waves. In this model, fast magnetoacoustic sausage-mode waves were suggested as the driver for the modulation of the background plasma emission. \citet{2013A&A...550A...1K} used this model to simulate fiber bursts in the dynamic spectrum and demonstrated that they can present negative frequency drifts with asymmetric emission-absorption features.

Of particular interest for fiber bursts is their potential for measuring the magnetic field strength in the corona, because all the major models suggest that the fiber drift rate and/or the emission-absorption frequency separation depend on the local magnetic field strength. However, the whistler wave model and the Alfv\'en wave model often predict very different magnetic fields in the corona. For example, \citet{1998A&A...333.1034B} summarized observations of dm-$\lambda$ fiber bursts at 1--3 GHz and used them to examine the existing models. They found that the Alfv\'en wave model predicts magnetic fields $\sim$4 times higher than the whistler wave model. They concluded, however, that it is difficult to validate the two models due to free parameters introduced without observational constraints, such as the magnetic field scale height and angle of the field with respect to the direction of the density gradient. Therefore, the key to unlocking their potential as a tool for diagnosing the coronal magnetic field lies in the identification and validation of the emission models.

However, the majority of fiber bursts studies are based on total-power dynamic spectral observations without simultaneous imaging data, which makes the identification of the exact exciter, and the responsible emission process by which fiber bursts are produced, particularly difficult. In one case, \citet{2005A&A...435.1137A} combined dynamic spectral data from a spectrograph (Astrophysical Institute Potsdam) and radio imaging data at a few separate frequencies from an interferometer (Nan\c{c}ay Radio Heliograph, or NRH) to study a fiber burst event in 200--600 MHz. By using a potential field extrapolation and assuming a coronal density model, they found the derived field strength using the whistler wave model \citep{1975SoPh...44..173K} is consistent with that given by the potential field extrapolation. The derived whistler group velocity is around 4,000 km$\cdot$s$^{-1}$, which is on average 24 times faster than the Alfv\'en speed they derived.

While the imaging constraints from NRH were important for the above conclusions, a single fiber burst in the dynamic spectrum was sampled at only two NRH frequencies for imaging. In order to trace of the propagation of fiber bursts continuously in time and frequency, simultaneous imaging observations are required to sample all times and frequencies where the bursts occur. Such an observing technique is known as ``dynamic imaging spectroscopy'', which has been applied to the study of dm-$\lambda$ type III bursts and stochastic spikes bursts very recently \citep{2013ApJ...763L..21C, 2015Sci...350.1238C}.

In the present paper, we report the first use of dynamic imaging spectroscopy with the Karl G. Jansky Very Large Array (VLA) to study dm-$\lambda$ fiber bursts during a prolonged eruptive solar flare. The VLA allows us to produce radio images with spectrograph-like time and frequency resolution (50 ms and 1 MHz, respectively) in a wide frequency range (1 GHz). Our study attempts to attack the following questions on dm-$\lambda$ fiber bursts: (1) What is the spatial relationship between the source producing the fiber bursts and the type IV continuum? (2) How are they associated with the flare energy release? (3) How and where does the fiber burst propagate in the corona? (4) Which theoretical model is the most consistent with the newly available dynamic imaging spectroscopic observations? In Section \ref{sec:overview}, we describe the instrumentation and present an overview of the event. Data analysis and results are presented in Section \ref{sec:fiber trajectories}. We compare observational results with fiber burst models in Section \ref{sec:theories}. We then conclude in Section \ref{sec:conclusions}.
\newline
\section{Overview of the Event} \label{sec:overview}
\begin{figure}[h]
\epsscale{0.9}
\figurenum{1}
\plotone{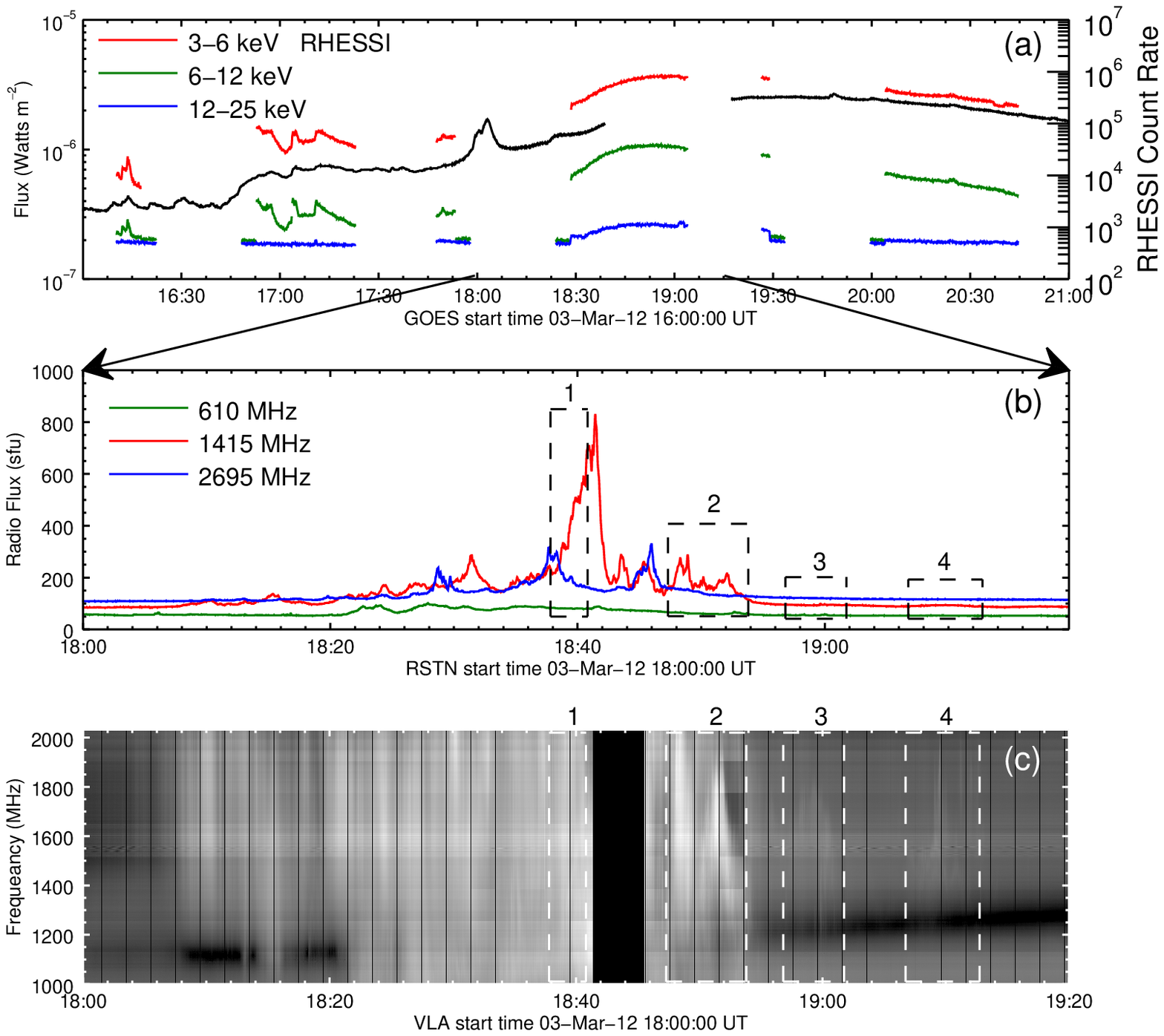}
\caption{(a) 5-hour GOES SXR 1--8{\AA} light curve (black solid line) covering the preflare phase, flare rise phase, and part of the decay phase. The HXR light curves of RHESSI energy bands 3--6 keV, 6--12 keV and 12--25 keV are plotted in red, green and blue color, respectively. (b) One-hour RSTN radio flux at 610 MHz (green solid line), 1415 MHz (red solid line) and 2695 MHz (blue solid line) starting from 18:00 UT. The dashed boxes highlight 4 regions of interest (ROI) with the fiber bursts --- see text. (c) The corresponding VLA cross-power dynamic spectrum is shown in the frequency range of 1--2 GHz. The solar data is temporally unavailable around 18:41--18:47 UT due to a calibrator scan.\label{fig1}}
\end{figure}

The dm-$\lambda$ fiber bursts in question were observed in an eruptive flare event in NOAA active region (AR) 11429 (N18E54) on 2012 March 3 (IAU solar flare identification SOL2012-03-03T17:00:04). A detailed description of this flare event is available from \citet{2014ApJ...794..149C}. Another study \citep{2015Sci...350.1238C} is devoted to the radio signature of a flare termination shock also observed in this event. To briefly recap, Figure~\ref{fig1}(a) shows a 5-hour plot of GOES (Geostationary Operational Environmental Satellite) soft X-ray (SXR; 1--8\AA) flux of the whole event starting from 16:00 UT, together with light curves from other X-ray energies. \citet{2014ApJ...794..149C} found supporting evidence that a fast white light coronal mass ejection (CME) and the long-duration C1.9 flare (peaked at 19:33 UT) are directly associated with an eruptive magnetic flux rope. Following the C1.2 SXR peak at 18:03 UT when the flux rope is quickly ejected from the lower corona, various types of dm-$\lambda$ fine structures, including pulsations, spike bursts, and fiber bursts are observed. Figure~\ref{fig1}(b) shows the radio flux record at the U.S. Air Force Radio Solar Telescope Network (RSTN) frequencies of 610 MHz, 1415 MHz, and 2695 MHz. Figure~\ref{fig1}(c) corresponds to an 80-min VLA cross-power dynamic spectrum at a short baseline. Between 18:10--19:00 UT, the dynamic spectrum shows a broadband, slowly-varying continuum emission, strongly polarized in left-hand circular polarization (LCP). It correlates well in time with the long rise in SXR emission. The dm-$\lambda$ fiber bursts are observed mainly in four regions of interest (ROIs) in Figure~\ref{fig1}(b - c). The fiber bursts are not visible at this reduced scale. Examples of fiber bursts in ROIs 1, 2 and 4 are shown in the full-resolution dynamic spectra of Figure~\ref{fig2}. The fibers are intermittent before 18:34 UT, weakly modulating the type IV continuum, while the broadband pulsations are the dominant fine structure in this period, as in ROI 1. They are more evident later in the event, and occur in several groups with each period lasting for more than 1 minute, as in ROI 2. The fibers become weaker after 18:55 UT when the continuum emission decays to the background level (Figure~\ref{fig1}(b)), but two major groups of fiber bursts are visible against the background as shown in ROIs 3--4.

%\hfill \break
%\newpage
\subsection{Instrumentation} \label{sec:intrumentation}
The primary instrument used in this study is the VLA, a general-purpose radio interferometer consisting of 27 antennas, each with a diameter of 25 m. These antennas can be reconfigured along three equiangular arms, with a maximum baseline $B_{\rm max}$ of 36 km. Recently, VLA has completed a major upgrade with significantly enhanced capabilities \citep{2011ApJ...739L...1P}. The ``solar observing mode'' was commissioned in 2011 \citep{2013ApJ...763L..21C}, which enables ultra-high cadence (up to tens of milliseconds) imaging spectroscopic observations of the Sun in frequency bands between 1--8 GHz, and will cover up to 50 GHz in the future. The high frequency and time resolutions of this new instrument, along with its full imaging capabilities, are ideal for studying fine structures of coherent solar radio bursts.

On 2012 March 3, the VLA was used to observe the full Sun in right- and left-circular polarizations in 1--2 GHz in its C configuration ($B_{\mathrm{max}}$ = 3.4 km). The resolution of the visibility data is 1 MHz in frequency and 50 ms in time, which is high enough to resolve the spectrotemporal features of fiber bursts. Due to data throughput limitations of the recording system at that time, when the array was still undergoing commissioning, only 15 antennas could be used for observing. The nominal angular resolution of the radio images scale linearly with frequency: at 2 GHz, the size of the synthesis beam is $\theta=14\arcsec\times10\arcsec$, and expands to $26\arcsec\times19\arcsec$ at 1 GHz. However, for radio sources with a simple morphology (which applies to our case), the location of the source centroid can be obtained with much higher accuracy, which scales inversely with the signal-to-noise ratio (SNR) of the radio image ($\sigma \approx \theta/\rm SNR$ \citet{1988ApJ...330..809R}). In our case, the SNR can easily reach above 20 for a bright fiber burst. In Section \ref{sec:fiber trajectories}, we will show that, based on the observation of a group of similar fiber bursts occurring in a short time window, uncertainties of the mean fiber source location can be further reduced.

To study the radio source in association with the flare energy release and the local magnetic field, we also obtained data from the following instruments:
\begin{itemize}
\setlength\itemsep{0em}
\item
Atmospheric Imaging Assembly (AIA) aboard NASA's Solar Dynamics Observatory (SDO) spacecraft: AIA obtains full-disk images with a nominal spatial resolution of $1.5\arcsec$, and a cadence of 10--12 s in multiple UV/EUV passbands from 1700 {\AA} to 94 {\AA}. These bands are sensitive to different temperatures in the solar atmosphere, from the chromosphere up to the solar corona \citep{2012SoPh..275...17L}.

\item
Reuven Ramaty High Energy Solar Spectroscopic Imager (RHESSI): RHESSI \citep{2002SoPh..210....3L} covers a broad energy range in X-rays and $\gamma$-rays from 3 keV to 17 MeV, and provides spectral imaging data with up to $2\arcsec$ spatial resolution, $\sim$1 keV spectral resolution, and up to 4 s temporal resolution.
\item
Helioseismic and Magnetic Imager (HMI) aboard SDO: HMI provides full-disk magnetogram data with a resolution of $1\arcsec$, which we use as the input to the Potential Field Source Surface (PFSS) model. The model is part of the Solar Software (SSW) package, which includes calculation of 3D global magnetic fields from the photosphere up to the source surface at $\sim$2.5$R_{\odot}$ \citep{2003SoPh..212..165S, 2008ApJ...673L.207N}.
\end{itemize}

\subsection{Dynamic Spectra of Fibers} \label{sec:dynamic spectra}
\begin{figure}[h!]
\epsscale{0.9}
\figurenum{2}
\plotone{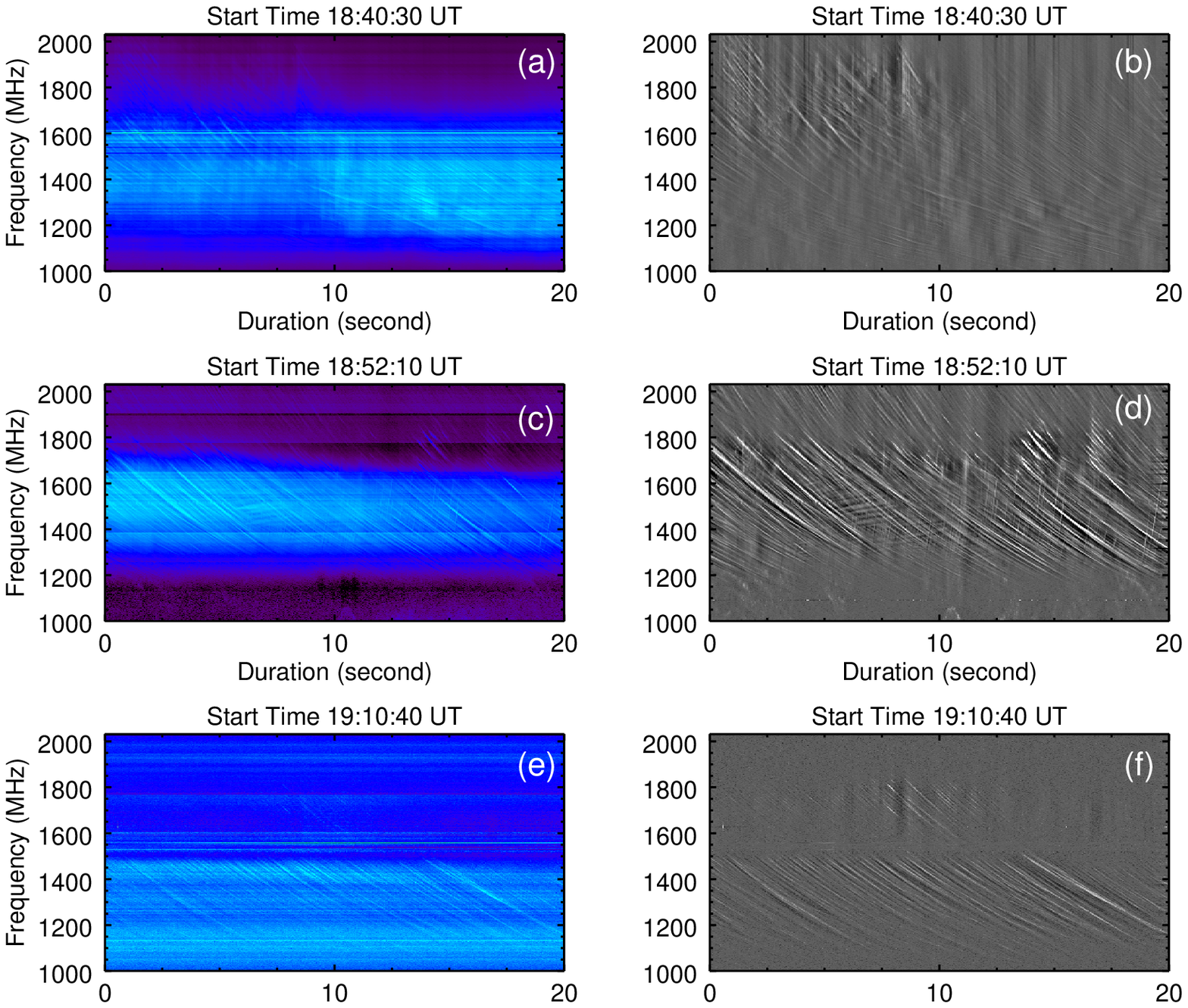}
\caption{Examples of dm--$\lambda$ fiber bursts in three regions of interest (ROI). Left panels ((a), (c) and (e)) show the original cross-power dynamic spectrum. Right panels ((b), (d) and (f)) show the corresponding high-pass filtered dynamic spectrum. Panels (a--b) show the fiber bursts in ROI 1 (marked as ``1'' in Figure~\ref{fig1}). Fibers with two apparently distinct drift rates exist in this period of time. Broadband pulsations (faint vertical features) coexist with the occurrence of fiber bursts (Panel (b)). Panels (c--d) show another detailed segment of fiber bursts in ROI 2. Emissions of Fiber bursts and continuum become stronger. Panels (e--f) show the decay phase of fiber bursts along with significantly reduced continuum emission and pulsations, corresponding to ROI 4. \label{fig2}}
\end{figure}

In Figure~\ref{fig2}, we show more detailed spectral features of the dm-$\lambda$ fiber bursts in three ROIs. The original cross-power dynamic spectrum and the high-pass filtered dynamic spectrum are shown in the left and right column, respectively. In the high-pass filtered dynamic spectrum, fast time-varying fine structures have been enhanced by scaling the original flux level relative to a smoothed background (by using a moving 3-s time window). Figure~\ref{fig2}(a--b) show detailed fine structures of fiber bursts in ROI 1, which seem to consist of two groups with different drift rates overlapping each other. For ROI 2 in Figure~\ref{fig2}(c--d), on top of the dominating fiber bursts with negative frequency drifts, it is interesting to note some narrow-band, fast-drift bursts with positive drifts (or ``reverse-slope'' fiber-like bursts) around the 10-s mark and later (detailed analysis of these peculiar bursts are outside the scope of the present work). For ROI 4 in Figure~\ref{fig2}(e--f), the upper bound of continuum emission has dropped to below 2 GHz. The fiber bursts in this episode are largely confined to a narrower frequency range, but their drift rates are similar to other groups of fiber bursts at earlier times. Statistics of the fiber drift rates will be presented in Section \ref{sec:fiber trajectories}.

\subsection{Fiber Images} \label{sec:fiber images}
\begin{figure}[h]
\epsscale{1}
\figurenum{3}
\plotone{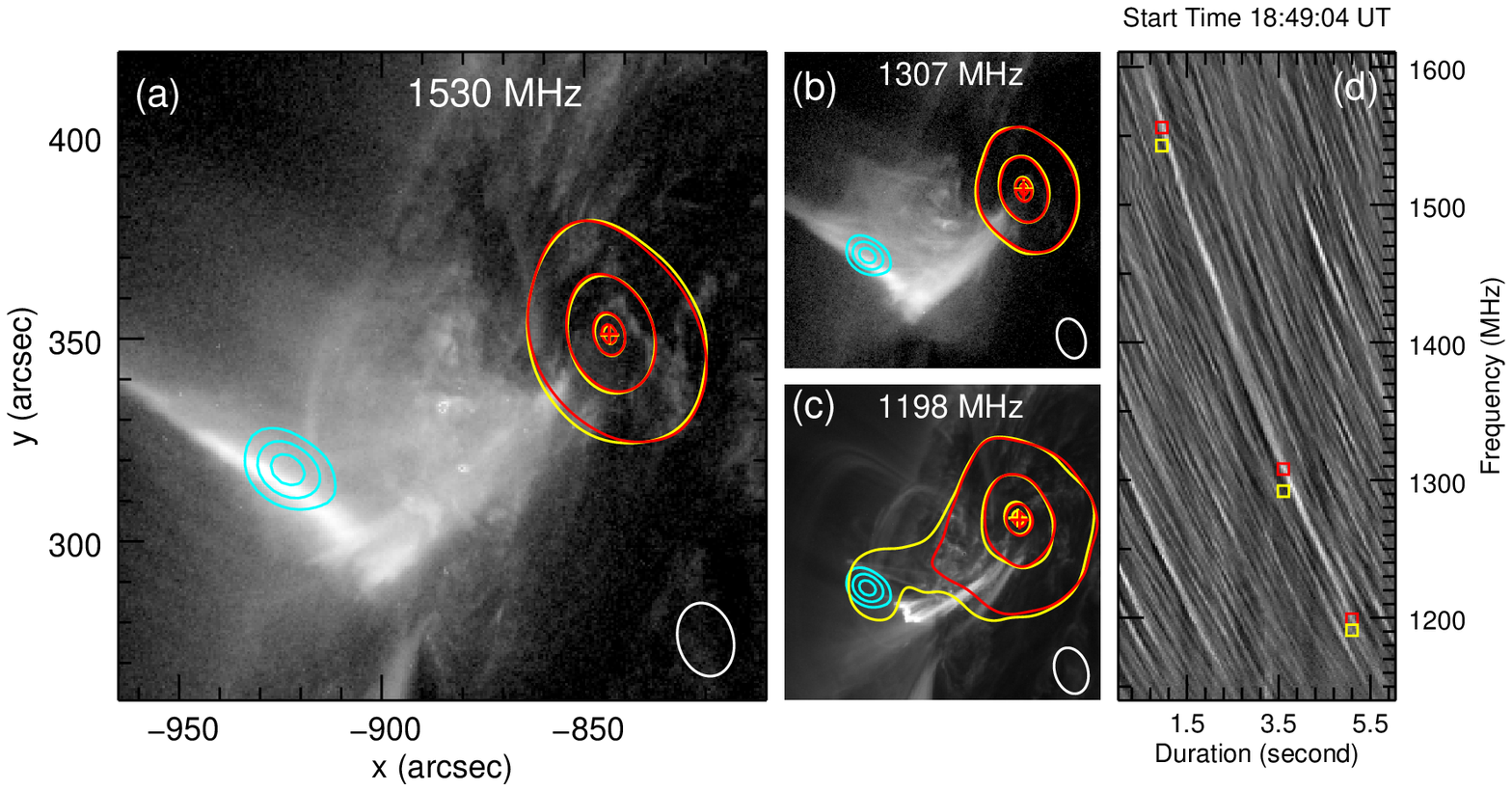}
\caption{(a--c) Three examples showing the morphology of the 12--25 keV HXR source (cyan), and the microwave source components (red and yellow) obtained from an individual bursts in the dynamic spectrum (d). The contour levels of the HXR source corresponds to 50\%, 70\%, 90\% of the maximum, and radio sources 30\%, 70\%, 95\% and 99\% of their peak values. Red contours of (a--c) correspond to the emission ridge of the fiber burst, also indicated by red boxes in (d). Yellow contours of (a--c) correspond to the adjacent absorption edge at the low frequency side of the corresponding emission ridge in (d). Around 18:50 UT when fiber bursts are very active, hot flare loops are visible only in the hot AIA passbands, as shown in the background images in (a--b), which correspond to AIA 131 {\AA} and 94 {\AA}, respectively. Around 19:28 UT, the coronal loops become evident in AIA 171 {\AA}, as shown in the background of (c). \label{fig3}}
\end{figure}

Figure~\ref{fig3}(a--c) show an example of VLA radio images sampled at three different times and frequencies along one well-defined bright fiber burst in the dynamic spectrum (Figure~\ref{fig3}(d)). At each time, we produce images on the emission ridge and the adjacent absorption edge at the same frequency (red and yellow square symbols in Figure~\ref{fig3}(d), respectively), which are shown in Figure~\ref{fig3}(a--c) as red and yellow contours overlaid on AIA EUV images.

Both sources in the emission and absorption edge are strongly polarized in LCP, and are located near the northwest footpoint of a coronal loop with a positive magnetic polarity, suggesting they are associated with o-mode plasma radiation. They show a Gaussian-like morphology with nearly indistinguishable shapes (Figure~\ref{fig3}(a--b)). At around 1190 MHz, the source obtained from the absorption edge develops a double-source configuration as shown by the yellow contours in Figure~\ref{fig3}(c): It shows another source is located above the limb, which coincides with a coronal hard X-ray (HXR) source observed by RHESSI at 12--25 keV (cyan contours) and hot flare loops seen by AIA 131 and 94 \AA\ passbands. \citet{2015Sci...350.1238C} identified this source as stochastic spike bursts, which appeared in the flare rise phase since 18:30 UT, and were interpreted as radio emissions from a flare termination shock driven by super-magnetosonic reconnection outflows from a reconnection site high in the corona.
%\hfill \break
We note that if the fiber emission is superposed on a background continuum, the radio intensity from the emission ridge should contain not only the contribution from the fiber burst itself, but also that from the background continuum. In order to evaluate the relation between the fiber emission and background, we produce radio maps of the ``net'' fiber source by subtracting the visibilities at the adjacent fiber absorption edge (as a proxy of the background continuum) from the emission ridge. We find the resulting net fiber sources have nearly identical morphology as those before the subtraction, with only differences in the absolute intensity. This is strong evidence that the fiber source and background continuum share the same origin. More likely, the observed fibers are modulations of the background emission. In the following analysis, for simplicity, we will focus on the source at the emission ridge and refer it to as the ``fiber source.''

The observed fiber and continuum sources are considerably larger than the size of the synthesized beam, whose half-power width is indicated by the white oval on the bottom-right corner of each image. The observed radio sources are thus spatially resolved. To obtain the size and shape of the ``actual'' source, we further deconvolve the radio images using the synthesized beam. The deconvolved fiber source retains a Gaussian shape, with a size of about $47\arcsec \times 39\arcsec$ at 1.4 GHz. The brightness temperature of the total source (fiber and continuum) can thus be estimated to be $6.9\times10^7$~K, and the corresponding net fiber component (after background subtraction) generally reaches 30\%--40\% of the total brightness temperature, or 2--$3\times10^7$~K. We note that the size of the deconvolved fiber images may still be significantly larger than the intrinsic value because of angular broadening due to scattering by inhomogeneous structures in the corona, which could itself amount to a few times 10 arcsecs near the limb at 1 GHz \citep{1994ApJ...426..774B}. Thus the apparent source size should be considered as an upper limit and accordingly, the brightness temperature estimate given above should be a lower limit.

\section{Spatially-Resolved Fiber Trajectories} \label{sec:fiber trajectories}
To perform detailed analysis of the observed thousands of fiber bursts with dynamic imaging spectroscopy, we first need to identify individual fibers in the radio dynamic spectrum. We adopted the code developed by \citet{2010SoPh..262..235A}, which is customized for tracing coronal loops in EUV and SXR images. With some modification, this algorithm can be applied to track fiber bursts in the dynamic spectrum with adequate contrast against the background. Details about this algorithm are already described in \citet{2010SoPh..262..235A}. Here we briefly summarize the main steps: (1) Locate the brightest pixel in the image, and perform an angular integration about the pixel center; (2) Determine the direction of the maximal integral, and use the neighboring pixel along that direction as a new tracing center; (3) Repeat the previous step and find all the subsequent time- and frequency-pixels along a fiber burst, and subtract them from the dynamic spectrum; (4) loop over the steps (1)--(3) until the residual is smaller than a pre-defined threshold. The start and stop frequency of any individual fiber depends on several factors, such as the frequency extent of the underlying continuum, the extent to which they may approach and even cross other fibers, and their intensity contrast.  In addition, the algorithm does not handle well the occasional frequency gaps due to radio frequency interference and fluctuations in contrast.  Hence, some of the shorter fibers we trace may be parts of longer fibers, but this should not affect our statistics.

\begin{figure}[h]
\epsscale{1.0}
\figurenum{4}
\plotone{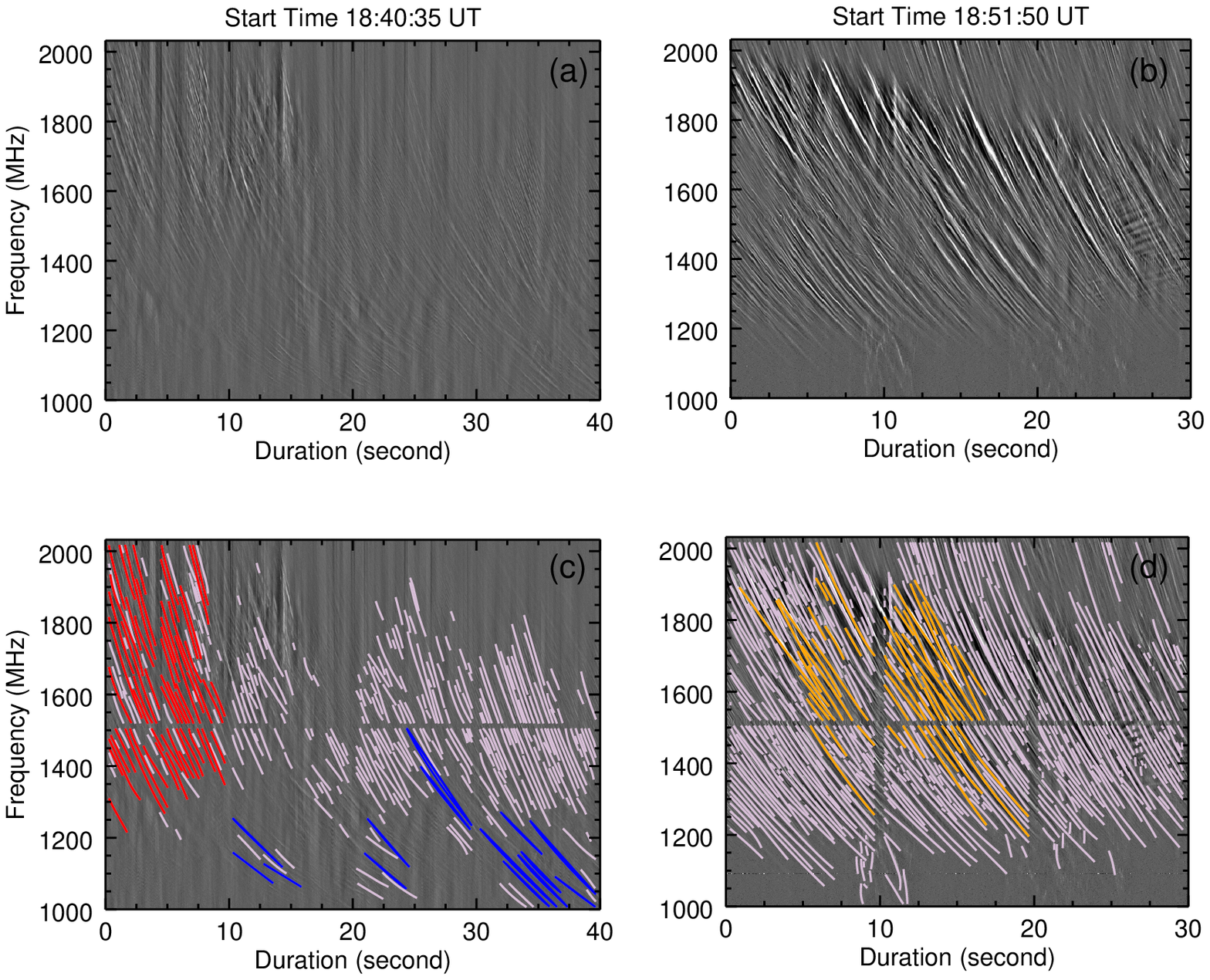}
\caption{(a--b) Two examples of high-pass dynamic spectrum from ROI 1 and ROI 2. The resolution is 50 ms in time and 1 MHz in frequency. The automatic tracing was done in separate 10-s segments. Results are shown in (c--d), with solid lines in light gray representing the overall detected line structures. The tracing procedures are interrupted around 1.5 GHz due to the presence of radio frequency interference (RFI). To obtain the frequency-drift fiber burst centroids, we select some particular fibers with similar drift rates, and highlight them in different color in (c) and (d). The two distinct drift families in (c) are indicated in blue and red, corresponding to the slow- and fast-drift fibers, respectively. The abnormally ``slow'' fibers only appear in ROI 1, which are confined to a narrower frequency range around 1.0--1.4 GHz than the commonly-observed ``fast'' fibers. The fiber bursts in (d) occur in the decay phase of the Type IV continuum emissions. They (in orange) are also used for comparison in the following studies.
\label{fig4}}
\end{figure}

Figure~\ref{fig4}(a--b) show two examples of dm-$\lambda$ fiber bursts corresponding to ROI 1 containing two groups with distinct drift properties, and another in ROI 2 during the decay phase of type IV continuum emissions. Their tracing results are compared in Figure~\ref{fig4}(c--d). The detection performs well in each ROI, with a total of 496 and 830 fiber bursts detected by the algorithm, as indicated by gray solid lines in each figure.

We next perform dynamic spectroscopic imaging to derive the trajectory of each identified fiber burst in the flaring region using the following procedure: (1) For each time integration, the frequency of the local emission maxima of each fiber burst is obtained using the tracing algorithm described above; (2) A radio image is produced for each identified frequency-time pair along the fiber burst in the dynamic spectrum; (3) The centroid position of each image is obtained by using a parabolic fit on several pixels around the image maxima. As already described in Section \ref{sec:intrumentation}, the uncertainty of the centroid position is estimated by using $\sigma \approx \theta_{\rm beam}/\rm SNR$, where SNR is the ratio of the image maximum to the root-mean-square (rms) of the image background where no emission is present. This way, for each identified fiber burst in the dynamic spectrum (i.e., in the time- and frequency-space), we obtain a series of image centroid positions in heliocentric X- and Y-coordinates, which effectively manifest themselves as the spatial trajectory of the fiber burst in the flaring region. The same practice is repeated for all identified fiber bursts, resulting in a collection of more than 1000 such fiber trajectories. We further refine our results by selecting only centroids with SNR greater than 20. We also flagged the centroids at the edges of the spectral windows (where the bandpass calibration is not as robust due to low gain of instrument) and eliminated those below 1.1 GHz where an additional source is present nearby and influences our centroid finding (see, e.g., Figure~\ref{fig3}(c)).

\begin{figure}[h]
\epsscale{0.9}
\figurenum{5}
\plotone{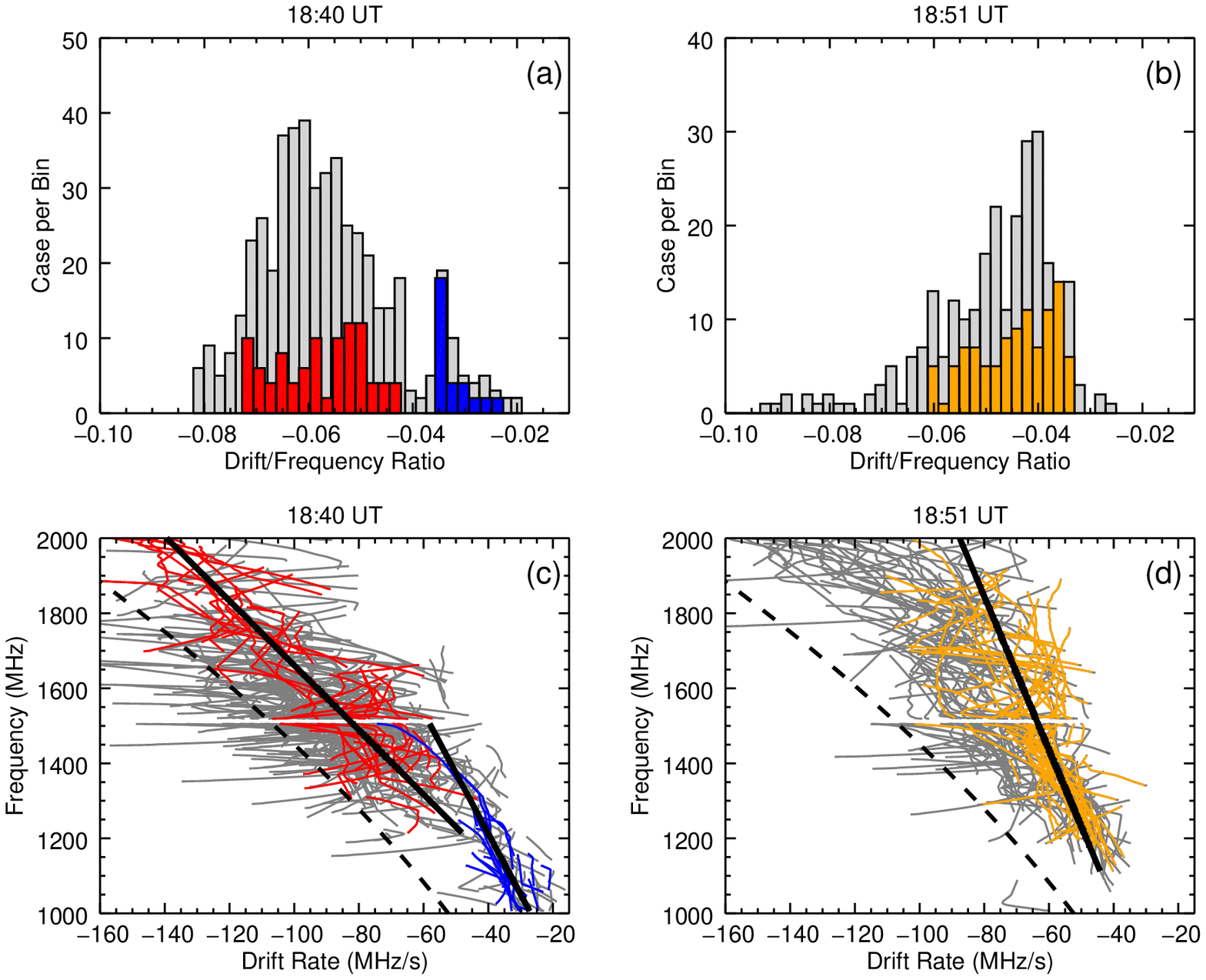}
\caption{(a--b) Histograms of normalized drift rates, $\dot{\nu}/\nu$, corresponding to the two periods in Figure \ref{fig4}. Each count represents one detected occurrence of drift burst. The color notation is the same as in Figure \ref{fig4}: the grey bars show the total distribution of detected fibers in each group, and the colored bars show the distribution of the colored fibers in Figure \ref{fig4}. (c--d) The corresponding frequency-dependent drift rates of the selected groups in 18:40 UT (blue and red), 18:51 UT (orange), and the total detected fibers (gray). The black solid lines indicate the mean drift rates of colored fibers sampled from each individual group. For comparison, the dashed curve is extrapolated from a second order approximation reported in \citet{1983PhDT.......135B} --- see text.\label{fig5}}
\end{figure}

We find that fiber bursts that appear together in the dynamic spectrum with similar frequency drift rates tend to have similar spatial trajectories, indicating that they propagate in the same loop system and share similar physical properties. However, they differ in their duration, spatial extent, and detailed curvature. To analyze their collective behaviors, we first identify groups of fiber bursts using their relative frequency drift rates $\dot{\nu}/\nu$ and then fit a mean trajectory to represent the group. Figure~\ref{fig5}(a--b) show the histogram of $\dot{\nu}/\nu$ based on the detected bursts in Figure~\ref{fig4}(c--d). The fibers in Figure \ref{fig5}(a) clearly show a bimodal distribution at around $\dot{\nu}/\nu\approx-0.06$ and $-0.038$, which we identify as a fast- and slow-drift fiber group respectively. In each fiber group, we only select the fibers that are within a short time window ($\lesssim 20$ s) to preserve their coherency, with drift rates near the peak of the distribution. For the right-skewed distribution of slow-fiber group, the range of $\dot{\nu}/\nu$ is selected between $-0.036$ and $-0.022$. For the distribution of two major groups (the fast-drift fibers at 18:40 UT and the fibers at 18:52 UT), a wider width ($\pm 0.013$) is used. The two subsets of slow- and fast-drift fibers are highlighted in blue (slow) and red (fast) in Figure~\ref{fig4}, and their distributions are plotted against the overall distribution, as indicated by colored bars in Figure~\ref{fig5}. The mean centroid trajectories of each of these fiber subsets will be determined and used in the following analysis. Another period at 18:52 UT shows a majority of drifts peaking around $\dot{\nu}/\nu\approx-0.047$. Similarly, we highlight the fiber subset ($\dot{\nu}/\nu$ between $-0.06$ and $-0.034$) in orange color. Lastly, for each traced burst in Figure~\ref{fig4}(c--d), we plot their frequency-dependent drift rates in Figure~\ref{fig5}(c--d). The mean drift rates are fitted with straight lines based on the sampled fibers. We will use them to derive the 3D source speed in Section \ref{sec:theories}. Overall, these drift rates are within the range reported for other observations in the frequency range 1--3 GHz \citep{1998A&A...333.1034B, 2006SoPh..236..155W}. The dashed curve plotted on Figure~\ref{fig5} (c--d) is a fit by \citet{1998A&A...333.1034B} to data by \citet{1983PhDT.......135B} taken near 200 MHz. Comparing with the \citet{1998A&A...333.1034B} study of 12 events, our drift rates rank among the lower of the drift rates they found.

\begin{figure}[h!]
\epsscale{0.95}
\figurenum{6}
\plotone{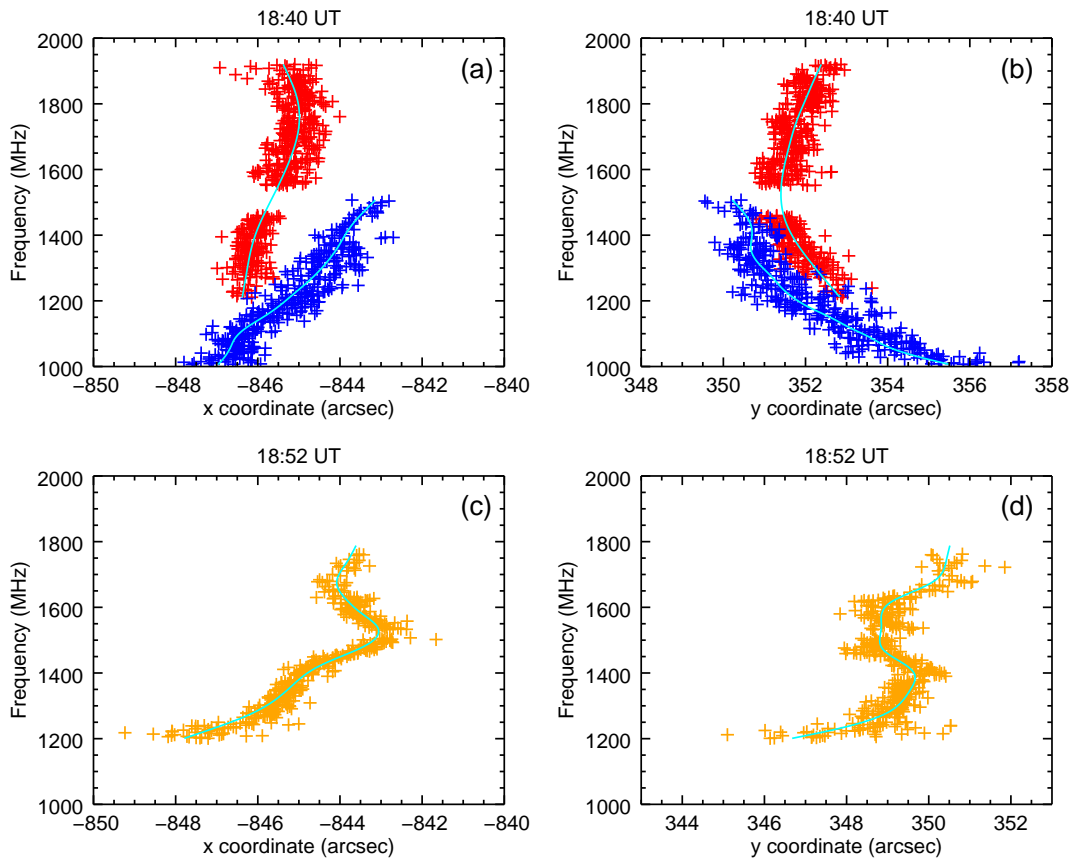}
\caption{The source centroid position of fiber bursts (color) as a function of frequency, obtained via the dynamic spectral imaging technique. (a) The x-component of burst centroid positions where the red and blue corresponds to the fast- and slow-drift group in 18:40 UT, respectively. (b) The same for y-component. (c--d) The same as (a--b), for the fibers in the period starting at 18:52 UT, where the fiber positions are shown in orange. The cyan solid lines indicate the mean fiber/type IV burst positions using smooth spline fitting with 5--8 degrees of freedom. \label{fig6}}
\end{figure}

Following this identification of fiber subsets, we used smooth spline functions with 5--8 degrees of freedom to fit to the x- and y-component of the two-dimensional mean trajectories respectively, as indicated by cyan solid curves in Figure~\ref{fig6}. Panels (a--b) show x- and y-centroid positions of the fast-drift (red) and slow-drift (blue) fibers around 18:40 UT as selected from the dynamic spectrum in Figure~\ref{fig4}(c). Panels (c--d) show another period around 18:52 UT (orange in Figure~\ref{fig4}(d)).

\begin{figure}[h!]
\epsscale{0.9}
\figurenum{7}
\plotone{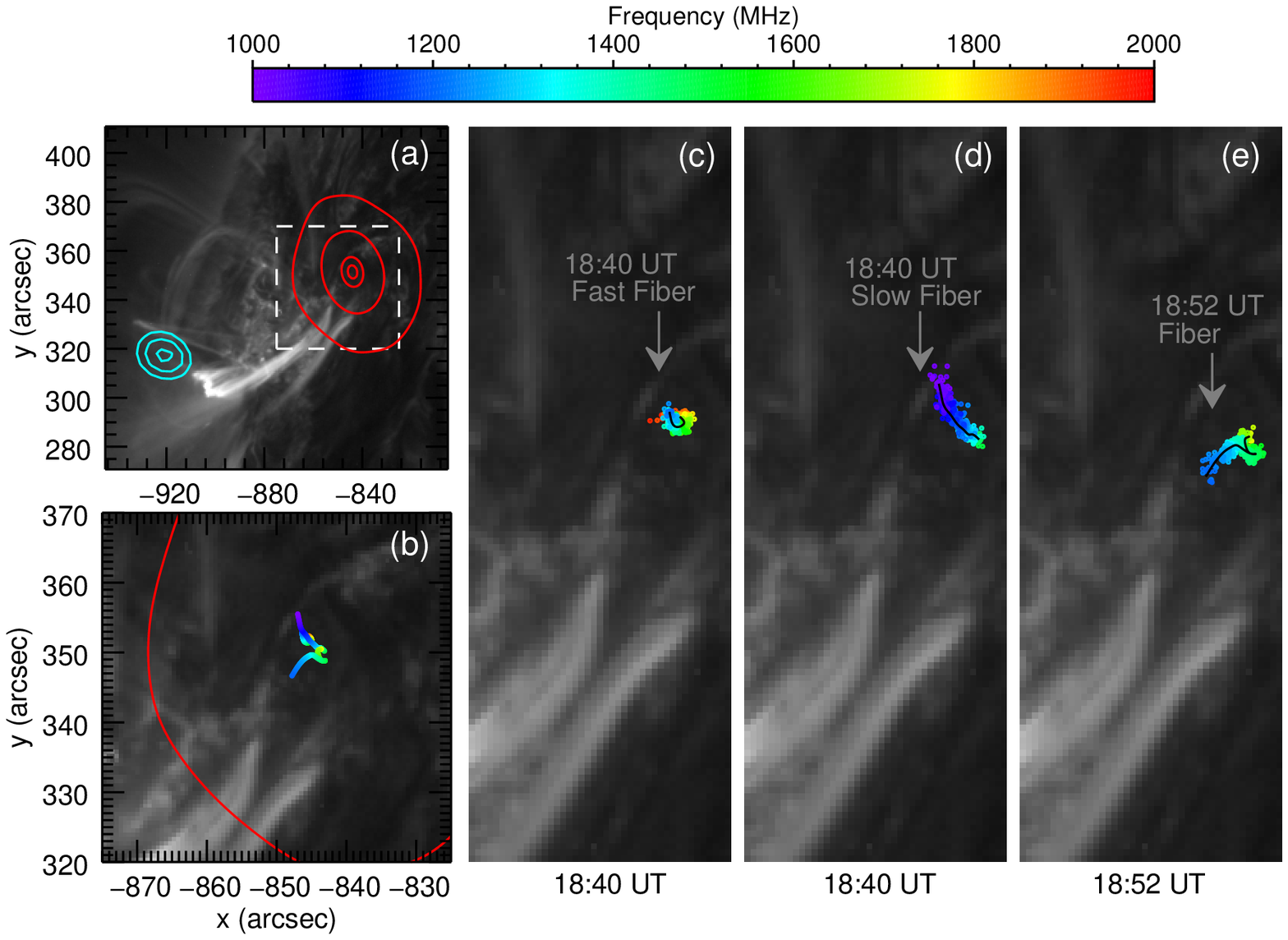}
\caption{(a) A comparison of the HXR and radio source components plotted over AIA 171 {\AA} at 19:28 UT. Their contour levels are similar to Figure~\ref{fig3}(b). (b) An overplot of all mean fiber burst trajectories, derived from 18:41 UT and 18:52 UT in Figure~\ref{fig6}. Colors from red to purple indicate different frequencies, as indicated by the color bar. The red contour corresponds to the outskirts of the fiber burst source, 30\% from each emission maximum. The actual measurements of fiber burst centroids in each region are plotted in (c--e), with black solid lines indicating the fitted trajectories.
\label{fig7}}
\end{figure}

Figure~\ref{fig7} shows direct comparison of fiber trajectories plotted over AIA 171 {\AA} at 19:28 UT. A snapshot of the source location, and the approximate size of HXR (cyan), fiber burst (red), are overplotted in panel (a). The white box marks the field of view (FOV) of the figure in panel (b). In panel (b), we compare all mean fiber trajectories derived from fiber subsets at 18:41 UT and 18:52 UT. The red contour show the 30\% level from their emission maximum. The centroids colored from red to purple indicate the trend towards lower frequencies (see color bar), generally equivalent to progressively greater heights in the corona. The individual centroid measurements are indicated by color dots overlaid with polynomial fits in black for each group separately in panels (c--e).
\hfill \break

\section{Theories and Discussion} \label{sec:theories}
In this section, we examine the implications of the new observational constraints from dynamic imaging spectroscopy for some leading fiber-burst models.  The observational implications of the models differ mainly in the propagation speed of the exciter or modulator, which in each case depends on the local magnetic field strength--hence, we will compare them in terms of their Alfv\'en Mach number $M_\mathrm{A}$. We assume, as do the models, that the fiber bursts emit near the local plasma frequency $\nu\approx \nu_{\mathrm{pe}}$:
\begin{equation}
\label{Eq_pe}
\nu_{\mathrm{pe}}\approx9\sqrt{n_\mathrm{e}[\mathrm{cm^{-3}}]}\ [\mathrm{kHz}].
\end{equation}
where $n_\mathrm{e}$ is the electron density. We adopt fundamental plasma radiation because the observed fiber bursts are highly polarized. Combined with a density model assumption, the frequency information of the fiber burst centroid can be used to estimate the source height in the corona. Here, we adopt a density model in the exponential form:
\begin{equation}
\label{Eq_dm}
n\left(r\right) =n_\mathrm{0}\ \mathrm{\exp}\left(-\frac{r}{r_\mathrm{n}}\right)
\end{equation}
where $n_\mathrm{0}$ is the reference value for density, $r_\mathrm{n}$ is the scale height of the exponential density model, and $r$ is the radial distance above the photosphere. The density is fixed by the frequency according to Eq. (\ref{Eq_pe}). By assuming some values of $n_{0}$ and $r_{n}$, the corresponding source height $r$ is deduced from Eq. (\ref{Eq_dm}). Together with the 2D centroid positions obtained from Section \ref{sec:fiber trajectories}, we can therefore reconstruct the density-model-dependent 3D centroid trajectories of the modulator as a function of frequency.

However, this leaves us with a density model with free parameters to be determined. In principle, knowing the modulator velocity from theory would provide the additional information needed to define the density model. In practice, we will derive the range of ``best-fit" density parameters for each competing fiber-burst model and discuss the implications.

On one hand, the true (deprojected) velocity of the fiber source can be directly derived from the 3D fiber centroid displacement and its drift rates:
\begin{equation}
\label{Eq_v3d}
V_{\mathrm{3D}}= -\frac{\mid\bigtriangleup\vec{S}\left(\nu\right)\mid}{\bigtriangleup\nu}\dot{\nu}\left(\nu\right)
\end{equation}
where $\bigtriangleup \vec{S}$ is the vector displacement of the modulator, and $\dot{\nu}$ is the frequency-dependent drift rate. Substituting Eq. (\ref{Eq_pe}) and Eq. (\ref{Eq_dm}), this suggests a simplified relationship:
\begin{equation}
\label{Eq_v3d2}
V_{\mathrm{3D}}\propto -\frac{\sqrt{n_{}}}{\nabla_{s}{n_{}}}{\dot{\nu}\left(\nu\right)}\approx -{r_\mathrm{n}}{\frac{\dot{\nu}\left(\nu\right)}{\nu}}
\end{equation}
where $\nabla_{s}{n_{}}$ is the density gradient along the fiber path, and we assume that the variation of projection angle along the propagating path is negligible. Eq. (\ref{Eq_v3d2}) shows that the deprojected velocity $V_{\mathrm{3D}}$ is explicitly dependent on the scale height $r_\mathrm{n}$ and the drift rate $\dot{\nu}$ \citep{1998A&A...333.1034B}.

On the other hand, the fiber-burst models predict different exciter speeds $V_{\mathrm{model}} \equiv M_\mathrm{A} v_\mathrm{A}$ along the reconstructed trajectory, where $M_\mathrm{A}$ is the Alfv\'en Mach number, and $v_\mathrm{A}$ is the Alfv\'en speed.  The Alfv\'en speed can be estimated by using the source height from the same assumed density model and the corresponding total magnetic field from the PFSS model. The earliest available PFSS model that includes AR 11429 is from 2012 Mar 5, when the region had rotated far enough onto the disk for its longitudinal fields to be measured by HMI. We use this model and rotate it back to the VLA observing time ($\sim$2 days earlier), with the unavoidable assumption that the AR underwent no significant evolution during that time. Although this model is based on the HMI magnetogram two days later, most features are similar to the actual AR configuration. Here, we only rely on the spatial distribution of the total magnetic field, while the detailed topology of AR magnetic field lines is of secondary importance.

Taking the magnetic field strength in the PFSS model at face value, both ways to deduce the velocity are constrained by two free parameters $n_\mathrm{0}$ and $r_\mathrm{n}$. In principle, we can find the optimum parameter-set iteratively. Briefly, the automatic fitting procedure is as follows: (i) select a trial value of density scale height $r_\mathrm{n}$, (ii) adjust $n_\mathrm{0}$ until a minimal difference between $V_{\mathrm{3D}}$ and $V_{\mathrm{model}}$ is obtained; (iii) decrease $r_\mathrm{n}$ if $V_{\mathrm{3D}} > V_{\mathrm{model}}$, and vice versa; (iii) repeat this until the optimum parameter-set is found.

\subsection{Whistler Wave Model} \label{sec:whistler}
The first type of theoretical model \citep{1975SoPh...44..173K, 1987SoPh..110..381M, 1990SoPh..130...75C} proposes that fiber bursts are due to the coalescence of whistler-mode waves and Langmuir waves in the corona, with the whistler frequency $\omega_\mathrm{w}\ll\omega_\mathrm{pe}$. This wave-wave coupling process produces transverse electromagnetic waves ($\omega_\mathrm{t}$) with $\omega_\mathrm{t}=\omega_\mathrm{w}+\omega_{\mathrm{pe}}$, inducing a flux enhancement on the emission ridge at $\omega_\mathrm{t}$, while depleting the local Langmuir waves ($\omega_{\mathrm{pe}}$) in the background type IV continuum to produce the absorption edge. Both wave modes could be excited by loss-cone instabilities of trapped electrons in the coronal loops. Therefore, this whistler wave model interprets individual drift bursts in the dynamic spectrum as a signature of whistler-mode wave packets traveling at the whistler group velocity through a stationary source region with enhanced Langmuir waves. In this interpretation, the frequency difference $\bigtriangleup\omega_{\mathrm{ea}}$ between the emission ridge and absorption edge is just the whistler frequency $\omega_\mathrm{w}$. When whistlers propagate parallel to the magnetic field, the group velocity $V_\mathrm{g}$ is given by \citet{1975SoPh...44..173K}:
\begin{equation}
\label{Eq_whistler}
V_\mathrm{g}=2c \frac{\omega_{\mathrm{ce}}}{\omega_{\mathrm{pe}}} \sqrt{x\left(1-x\right)^{3}} = M_{\mathrm{A}}^\mathrm{w} v_{\mathrm{A}}
\end{equation}
where $x$ denotes the frequency ratio in the following form:
\begin{equation}
\label{Eq_x}
x=\frac{\omega_\mathrm{w}}{\omega_{\mathrm{ce}}}=\frac{m_\mathrm{e}}{e} \frac{\omega_\mathrm{w}}{B}
\end{equation}
Here, $c$ is the speed of light, $\omega_{\mathrm{ce}}$ is the electron cyclotron frequency, $m_\mathrm{e}$ is mass of the electron, $e$ is the electron charge and $B$ is the total magnetic field. We further define an Alfv\'en Mach number for the whistler group speed $M_{\mathrm{A}}^\mathrm{w}=V_\mathrm{g}/V_{\mathrm{A}}$, which ranges from a few to a few tens under typical coronal conditions.

In Eq.~\ref{Eq_whistler}, the frequency of whistler waves is approximated by the mean emission-absorption separation: $\omega_\mathrm{w}\approx \bigtriangleup\bar\omega_{\mathrm{ea}}$. Among the three fiber groups, fast- and slow-drift groups in 18:40 UT, and the 18:52 UT group, $\bigtriangleup\bar\omega_{\mathrm{ea}}$ corresponds to 6.2, 8.11, and 6.73 MHz, respectively. Once the 3D trajectory is fixed by combining the statistically-fitted 2D centroid trajectory and the density model, $V_\mathrm{g}$ is calculated via Eq. (\ref{Eq_whistler}), using the derived values of $\omega_\mathrm{w}$, $\omega_{\mathrm{pe}}$ and $B$ along this trajectory.

\begin{figure}[b!]
\epsscale{0.93}
\figurenum{8}
\plotone{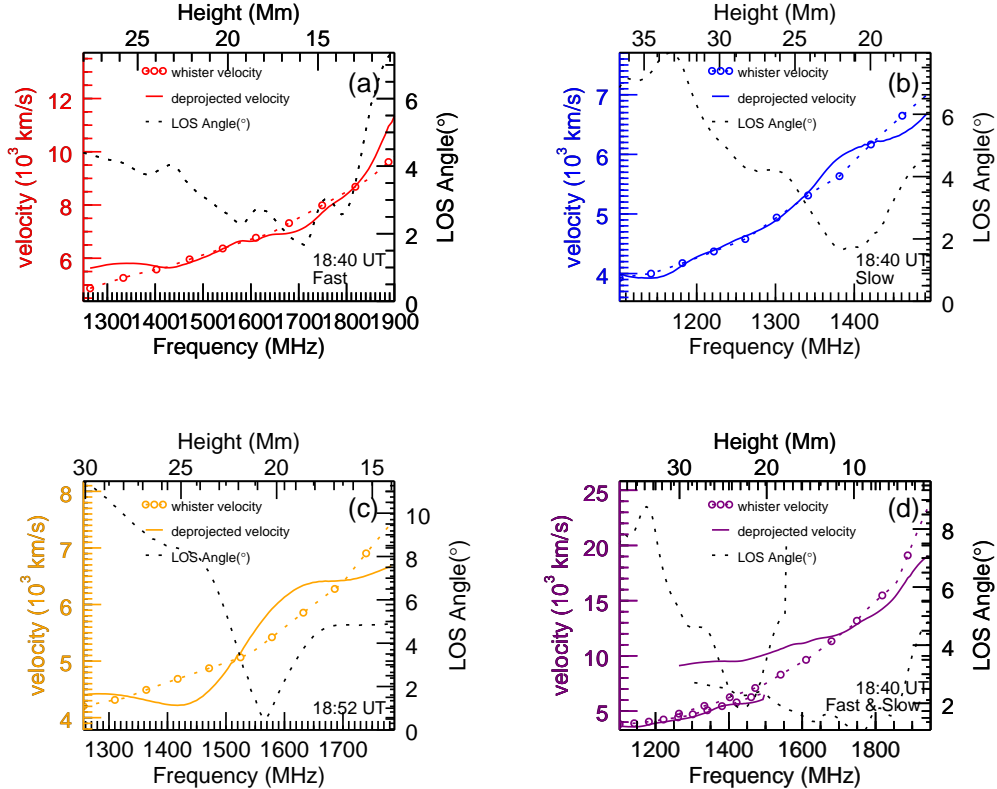}
\caption{Comparisons of the deprojected fiber burst source velocities derived from the observation (solid lines), and the whistler wave model (open circles) in three different cases. The LOS angles of the vector velocity are plotted in dash lines, in reference to the right vertical axis in each figure. (a) Results for the 18:40 UT fast-drift group, (b) 18:40 UT slow-drift group, and (c) 18:52 UT group, which were fit independently using different $n_\mathrm{0}$ and $r_\mathrm{n}$ in the density model. Panel (d) attempts to fit both fast- and slow-drift groups at 18:40 UT with a single density model. Table~\ref{tab1} provides a list of best-fit modeling parameters, and mean error percentage in each case. \label{fig8}}
\end{figure}

Next, we optimized the density model to achieve the best fit between $V_{\mathrm{3D}}$ (colored solid lines) and $V_\mathrm{g}$ (colored open circles), as shown in Figure~\ref{fig8}. Panels (a--b) show the results for the fast- and slow-drift groups at 18:40 UT. Panel (c) shows the results for the fiber group at 18:52 UT. In panel (d), we attempted to simultaneously fit the 18:40 UT fast- and slow-drift groups by using a single density model. In each plot, the LOS projection angle of $V_{\mathrm{3D}}$ is plotted as dotted black lines (referred to the right y-axis).  In all cases, this LOS projection angle is small ($<10^\circ$), meaning the propagation direction of the exciter is nearly along the LOS. The top x-axis in each plot indicates the radial source height above the photosphere, implicitly related to the frequency via the density model. Table~\ref{tab1} lists a summary of the corresponding parameters $n_\mathrm{0}$ and $r_\mathrm{n}$ for the best-fit density model, and quantitative results including the range of the density $n$ and magnetic field strength $B$ along each reconstructed trajectory. We also estimated the goodness of fit by calculating the mean fitting error along each trajectory, using $\mid\left(V_{\mathrm{3D}}-V_{\mathrm{model}}\right)/V_{\mathrm{model}}\mid$, where $V_{\mathrm{model}}$ represents the velocity estimated by the fiber burst model.

Based on these results, we have several findings: (i) in Figure~\ref{fig8}(a--c), each fiber group can be fit independently with $<10$\% mean fitting error. However, using a uniform density model to fit both drift types in 18:40 UT (Figure~\ref{fig8}(d)) significantly increases the fitting error to nearly 17\%, suggesting that the two drift-families of bursts are most consistent with being from different source regions; (ii) the fitting results from the two drift groups at 18:40 UT (c.f., Table \ref{tab1}) also suggest that fibers with observed faster drift rates can involve two coexisting factors: a steeper density gradient (smaller $r_\mathrm{n}$) and a faster propagating velocity. The fast-drift group has an average velocity of $V_\mathrm{g}$ = 7,000~km$\cdot s^{-1}$ and $r_\mathrm{n} = 1.97\times10^{10}$~cm, while the slow-drift group has $V_\mathrm{g}$ = 5,000~km$\cdot s^{-1}$ on average and $r_\mathrm{n} = 3.34\times10^{10}$~cm; (iii) in all three fiber groups, the whistler waves (dashed lines with open circles in Figure~\ref{fig8}) decelerate as they propagate upward into weaker field regions, with a wide range between 12,000 km$\cdot s^{-1}$ and 4,000 km$\cdot s^{-1}$. The whistler group velocity is generally at least ten times faster than the local Alfv\'en speed (see $M_{\mathrm{A}}$ for the whistler model in Table~\ref{tab1}). This is consistent with the result reported by \citet{2005A&A...435.1137A}; (iv) in the frequency range of 1--2 GHz, where the fiber bursts occur, the density is typically in the order of $10^{10}$ cm$^{-3}$ over 10--36 Mm.

\begin{figure}[h]
\epsscale{1.0}
\figurenum{9}
\plotone{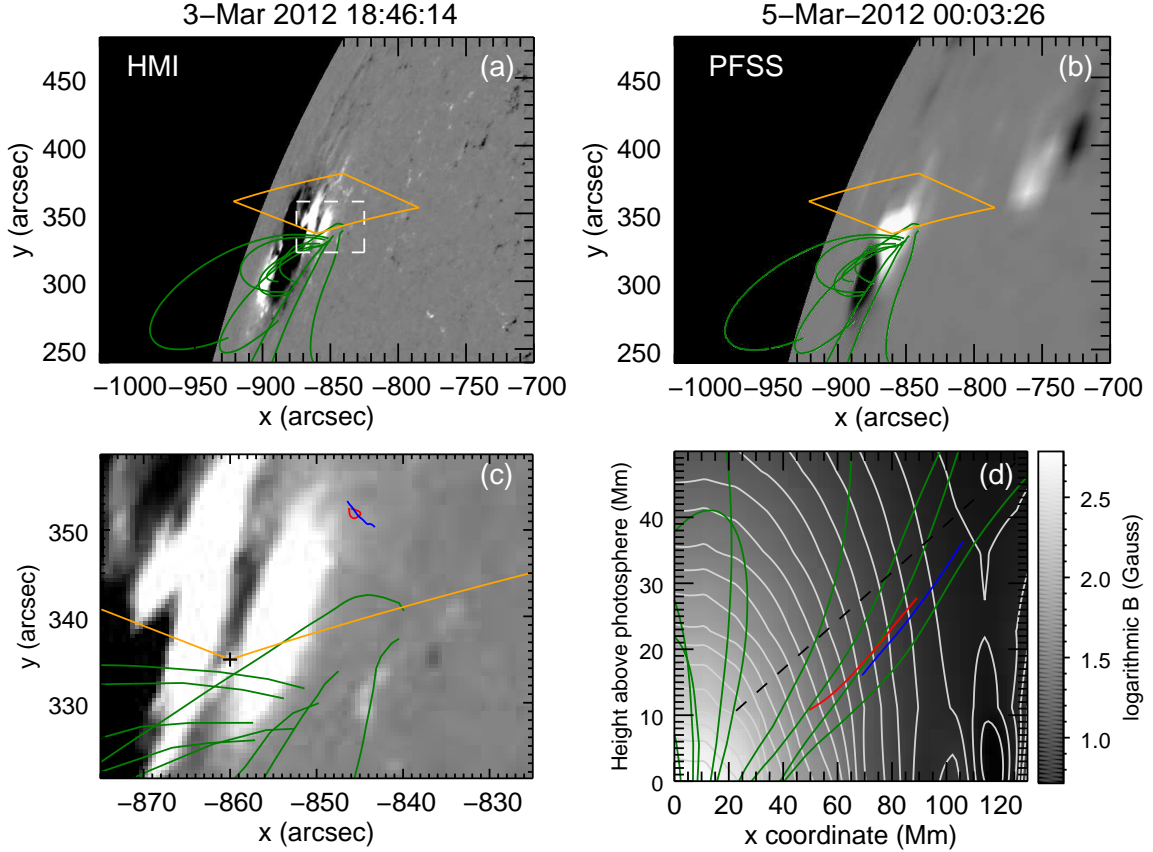}
\caption{(a) The HMI LOS magnetogram of AR 11429 around the VLA observing time. The green lines indicate selected field lines of the PFSS model. The orange lines indicate the boundary of the x vs. height plane used for projection (see text for details). (b) Same as (a), but showing the total magnetic field $B_\mathrm{t}$ at the base of PFSS model. (c) Zoomed view of the region outlined by the white dashed lines in (a), overplotted with mean-fit fiber burst trajectories of fast-drift group (red) and slow-drift group (blue) at 18:40 UT. (d) The background shows the distribution of total magnetic field $B_\mathrm{t}$ in logarithmic scale within the cross-section area, bounded by the orange lines in (a--c). Overplotted are the corresponding fiber trajectories in panel (c), and PFSS field lines (green) projected onto this plane. The black dash line indicates the projected line-of-sight, which is marked as a black cross symbol in panel (c).
\label{fig9}}
\end{figure}

Finally, we show heliocentric and cross-section perspectives of fiber trajectories based on the whistler fitting result. Figure~\ref{fig9}(a)\&(b) show an overview of some selected PFSS field lines (green) within or close to the AR 11429, with the background corresponding to the HMI LOS magnetogram Fig.~\ref{fig9}(a) and the total magnetic field at the base of PFSS model Fig.~\ref{fig9}(b). The orange boundary outlines a radially-vertical screen for projection in Figure~\ref{fig9}(d). It has an origin at (-860\arcsec, 335\arcsec), covering 130~Mm in the longitudinal direction, and $50$~Mm in the radial direction. Figure~\ref{fig9}(c) compares the two fiber trajectories at 18:40 UT from the heliocentric perspective, with the red and blue color indicating the fast-drift and slow-drift groups respectively. The 3D reconstructed fiber trajectories and PFSS field lines were split into their longitudinal and radial components relative to the projection screen, which allows us to compare their relative locations from the vertical screen perspective, as shown in Figure~\ref{fig9}(d). Both fiber trajectories, although they are not precisely parallel, follow a similar trend as the set of PFSS field lines (green) selected from Figure~\ref{fig9}(a). The theoretical expectation \citep{1975SoPh...44..173K} is that the whistler waves travel along the magnetic field, due both to refraction toward the field during propagation and enhanced Landau damping for any waves propagating at an appreciable angle to the field. The small angular discrepancy indicated in Figure~\ref{fig9}(d) may thus not be real, but may arise from uncertainties in the PFSS model, which was based on measurements 2 days after the event. From our 3D reconstruction, it becomes obvious that the apparent ``kink'' in the trajectory in Figure~\ref{fig9}(a) is a projection effect due to the motion being close to the line of sight.

%\hfill \break
\newpage
\subsection{MHD Wave Models} \label{sec:MHD}
\paragraph{Alfv\'en wave model} The earliest MHD model, often known as the `Alfv\'en wave model', considers the fiber bursts as a modulation of plasma emission by nonlinear solitons propagating with slightly super-Alfv\'enic velocity. When solitons enter the source region that is producing continuum type IV emission, such waves induce MHD disturbances in the local plasma parameters, followed by redistribution of the frequency of the emission. \citet{1990A&A...236..242T} showed that emission-absorption spectral properties can be formed by a compressional (dilutive) soliton: increased (decreased) density within the soliton shifts the radiation to higher (lower) frequencies (causing the emission ridge), which in turn reduces the emission at frequencies corresponding to the original undisturbed density at the soliton site (causing the absorption edge). Since the observed drift rates are an order of magnitude faster than the shock-generated type II bursts, the propagation speed $V_\mathrm{s}$ of a super-Alfv\'enic soliton is estimated between 1 to 3 times of the local Alfv\'en wave $v_{\mathrm{A}}$, given in the following form:
\begin{equation}
\label{Eq_soliton}
V_\mathrm{s}=M_{\mathrm{A}} v_{\mathrm{A}},
\end{equation}
where
\begin{equation}
\label{Eq_alfven}
v_{\mathrm{A}}=\frac{B}{\sqrt{4\pi n_{i} m_{i}}},
\end{equation}
$M_\mathrm{A}$ is the Mach number, $B$ is total magnetic field, and $n_\mathrm{i}\approx n_\mathrm{e}$ is the ion density. The mean ion mass $m_\mathrm{i}$ is approximated by 1.26 $m_\mathrm{p}$, the proton mass. At reconstructed source sites, $v_{\mathrm{A}}$ is derived from $n_\mathrm{e}$ and $B$, provided by the observed fiber frequency and PFSS model respectively.

\paragraph{Fast sausage magnetoacoustic wave model} An alternative MHD wave model that we examine considers the modulation of broadband type IV burst radiation by magnetoacoustic wave trains in a sausage mode \citep{2006SoPh..237..153K, 2013A&A...550A...1K}, which could be impulsively triggered by the flare. As mentioned above, the key challenge for the Alfv\'en wave model is that the total emission should be conserved, which is often inconsistent with observations. In contrast, sausage mode oscillations induce fluctuations of both plasma density and magnetic field strength as the modulator propagates, which can alter the strength of the emission. Using this MHD model, recent work by \citet{2006SoPh..237..153K} and \citet{2013A&A...550A...1K} have successfully simulated the observed emission-absorption asymmetry in the fiber burst spectrum. The phase velocity $v_{\mathrm{ph}}$ of such propagating waves is intermediate between the Alfv\'enic speeds inside and outside the magnetic tube, $v_{\mathrm{A_{in}}}\leqslant v_{\mathrm{ph}}\leqslant v_{\mathrm{A_{out}}}$. In the low-$\beta$ corona, pressure balance implies that the internal and external magnetic field strengths are approximately equal ($B_{\mathrm{in}}\approx B_{\mathrm{out}}$). Therefore, the phase velocity ratio $v_{\mathrm{A_{out}}}/v_{\mathrm{A_{in}}}$ is approximately equal to the square root of the density compression ratio inside and outside the flux tube $\sqrt{n_{\mathrm{in}}/n_{\mathrm{out}}}$. In this study, the modulator velocity $V_{\mathrm{sausage}}$ is defined as the following:
\begin{equation}
\label{Eq_sausage}
V_{\mathrm{sausage}} \equiv M_\mathrm{{A}}^\mathrm{{s}} v_\mathrm{A_{in}} < v_\mathrm{A_{out}} \approx \sqrt{n_\mathrm{in}/n_\mathrm{out}}v_\mathrm{A_{in}},
\end{equation}
where $M_\mathrm{A}^\mathrm{s}$ is the generalized Mach speed of the sausage mode wave with regard to $v_\mathrm{A_{in}}$.

To fit our MHD model to the data, we assume a reasonable range of Mach numbers in Eq.~\ref{Eq_soliton} and Eq.~\ref{Eq_sausage}. Here, we use $M_\mathrm{A}=1,2,3$ for both Alfv\'en wave model \citep{1990A&A...236..242T} and fast sausage magnetoacoustic wave model \citep{2003ApJ...598.1375A}. Under extreme conditions, $M_\mathrm{A}=10$ is possible for the fast sausage magnetoacoustic wave model. Such a high density ratio ($n_\mathrm{in}/n_\mathrm{out}\gtrsim 10^2$) for flare or postflare loops could be due to chromospheric evaporation \citep{2004ApJ...600..458A}.

\begin{figure}[h]
\epsscale{1.15}
\figurenum{10}
\plotone{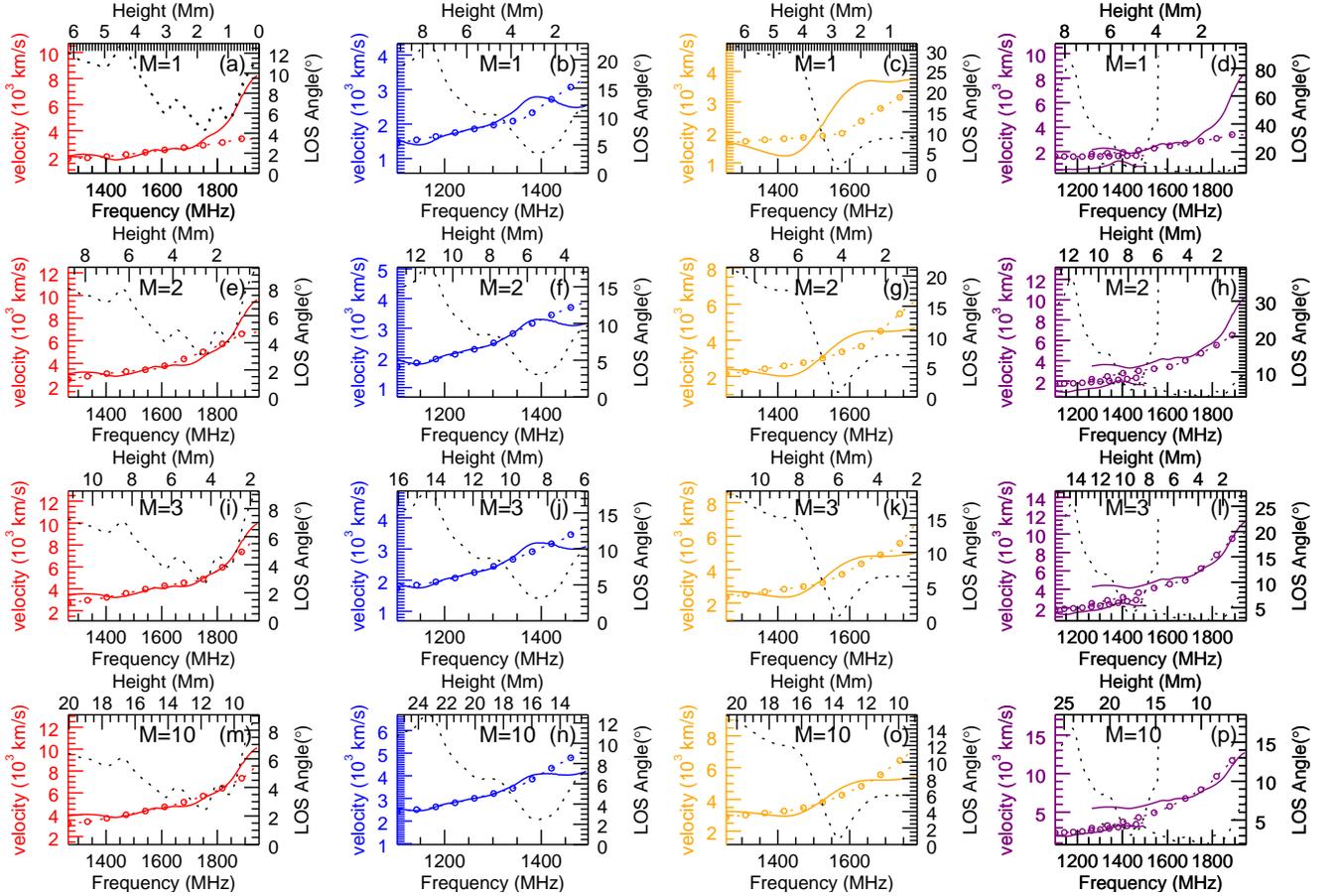}
\caption{Similar to Figure \ref{fig8}, the optimized fitting results between the observation (solid lines) and the MHD wave model (open circles). Columns starting from the left correspond to 18:40 UT fast-drift group (red), 18:40 UT slow-drift group (blue), 18:52 UT group (yellow), and lastly combined fitting of fast- and slow-drift group at 18:40 UT (purple), respectively. Rows starting from the top correspond to different assumption on Mach number $M_\mathrm{A}$ = 1, 2, 3 and 10 respectively. The corresponding density model parameters and fitting errors are also listed in Table~\ref{tab1}.\label{fig10}}
\end{figure}

Figure~\ref{fig10} shows the results based on the MHD models by using the fitting procedures described in Section \ref{sec:whistler}. The left-column panels from top to bottom correspond to the 18:40 UT fast-drift group when $M_\mathrm{A}=1, 2, 3, 10$, respectively. The other columns show the same for other cases: The 18:40 UT slow-drift group in the second column, the 18:52 UT group in the third column, and the single density model fit to both fast and slow drift groups at 18:40 UT in the last column. Again, fitting both drift-groups at 18:40 UT with a single density model (panel in the right-most column) fails to perform as well as separate density models. As before, the slow drift group can only be fit with a lower density gradient (larger $r_n$) than other groups. Similar quantitative comparisons are listed in Table~\ref{tab1}.

For a low Mach number (e.g. $M_\mathrm{A}=1, 2$), fiber trajectories are fitted towards unphysically low height (0--0.5~Mm above the photosphere) at high frequencies. The corresponding fitting errors are mainly due the discrepancies between slow MHD wave velocities $V_{\mathrm{MHD}}$ (open circles in Figure~\ref{fig10}) and the faster $V_{\mathrm{3D}}$ (solid lines in Figure~\ref{fig10}) at frequencies above 1.8~GHz, suggesting that the magnetic field strength, even at such a low height, is too weak to drive the fiber exciter at a speed comparable to $V_{\mathrm{3D}}$. The fitting becomes much easier with $M_\mathrm{A} =3, 10$. For $M_\mathrm{A} = 3$, the errors are reduced to $\lesssim 10\%$ in all cases, with a reasonable source height between 1.5 and 16~Mm, and the corresponding magnetic field strength B$\approx$ 290--30 Gauss. In the extreme case where $M_\mathrm{A} = 10$, the errors are only slightly reduced by 1--2$\%$. And the trajectories are fit towards higher heights in the range of 8--26~Mm where the magnetic field strengths are even weaker (85--13~G). These fitting parameters are comparable to the whistler result since the MHD wave velocity at 10 times of the local Alfv\'en speed is in fact similar to the whistler group speed.

Although the fit is less good for the MHD models than for the whistler model, the MHD wave models attempt to solve other serious shortcomings for the whistler model, especially the difficulty in accounting for the brightness of the emission, which requires some mechanism for maintaining the coherency of the waves over a large coronal volume.  We acknowledge that there are significant uncertainties in our use of the PFSS model, since the active region from which the extrapolation is performed has evolved over the 2 days between the event time and the measurements.
\hfill \break

\floattable
\begin{deluxetable}{cclccccc}
\tablecaption{List of parameters in whistler wave model and MHD wave model \label{tab1}}
\tablecolumns{7}
\tablenum{1}
\tablewidth{0pt}
\tablehead{
\colhead{Model} & \colhead{$M_\mathrm{A}$} & \colhead{ROI} & \colhead{$n_\mathrm{0}$} & \colhead{$r_\mathrm{n}$} & \colhead{$H$} & \colhead{$B$} & \colhead{Error} \\
&               &                & \colhead{($10^{10}$ cm$^{-3}$)} & \colhead{($10^{9}$ cm)} & \colhead{(Mm)} & \colhead{(Gauss)} & \colhead{($\%$)}
}

\startdata
   & 17--28 & fast-drift 18:40 UT & 8.06 & 1.97 & 10.7--27.8 & 62--11 & 5.5 \\
Whistler & 22--27 & slow-drift 18:40 UT & 4.45 & 3.34 & 15.9--36.3 & 24--8 & 1.9 \\
Model    & 20--28 & 18:52 UT            & 7.20 & 2.28 & 13.7--29.9 & 34--10 & 5.8 \\
\cline{1-8}
         & 1      & fast-drift 18:40 UT & 4.69 & 0.71 & 0--6.2 & 333-118 & 24.7 \\
         & 1      & slow-drift 18:40 UT & 2.85 & 1.42 & 0.4--9.2 & 251--80 & 8.5 \\
         & 1      & 18:52 UT            & 3.95 & 0.94 & 0--6.6 & 307--105 & 26.8 \\
\cline{2-8}
          & 2      & fast-drift 18:40 UT & 4.69 & 1.01 & 0--8.8 & 332--82 & 10.3 \\
          & 2      & slow-drift 18:40 UT & 3.24 & 1.69 & 2.7--13.0 & 147--47 & 4.5 \\
MHD      & 2      & 18:52 UT            & 4.10 & 1.25 & 0.5--9.3 & 279--68 & 13.4 \\
\cline{2-8}
Model    & 3      & fast-drift 18:40 UT & 5.36 & 1.13 & 1.5--11.2 & 288--58 & 7.9 \\
          & 3      & slow-drift 18:40 UT & 3.89 & 1.69 & 5.7--16.1 & 94--32 & 4.6 \\
          & 3      & 18:52 UT            & 4.53 & 1.39 & 1.9--11.7 & 201--49 & 10.2 \\
\cline{2-8}
          & 10     & fast-drift 18:40 UT & 8.77 & 1.34 & 8.4--20.1 & 85--20 & 7.9 \\
          & 10     & slow-drift 18:40 UT & 4.73 & 2.22 & 11.9--25.5 & 39--13 & 4.3 \\
          & 10     & 18:52 UT            & 6.63 & 1.69 & 8.8--20.7 & 65--18 & 8.5 \\
\enddata

\end{deluxetable}
\newpage
\section{Conclusions} \label{sec:conclusions}

In this study, we explored radio diagnostics of dm-$\lambda$ fiber bursts using the new technique of radio imaging spectroscopy enabled by the recently upgraded VLA. We also developed a framework of data analysis that can be used in the analysis of future observations. Major results are concluded in the following.

By combining multi-wavelength observations, we investigated the spatial association of the radio source with the EUV coronal loop structures and the HXR source. The fiber source is located close to the footpoint of the hot post-flare loops visible in AIA passbands that are sensitive to hot plasma temperatures (131, 94, and 335~{\AA}). A HXR loop-top source with a non-thermal power-law component is seen in the RHESSI data, suggesting the presence of flare electrons during the gradual rise of GOES SXR light curves. One possible interpretation of the lack of HXR footpoint sources is the dominance of trapping over precipitation of high energy electrons in the converging magnetic fields. It is likely that they are trapped in a magnetic mirror configuration with loss-cone type distribution, which, in turn, enhances the growth of localized plasma waves for producing the fiber bursts and the continuum decimetric emission. The centroid position and shape of the sources at the fiber emission ridge and at the absorption edge are very similar to each other, suggesting that the fiber burst and background continuum reside in the same loop system and share the same origin. The fiber bursts are strongly polarized in the sense of o-mode waves. The deconvolved fiber source size appears fairly large, typically about $47\arcsec \times 39\arcsec$ at 1.4 GHz. An estimate of the lower limit of the brightness temperature of the total on-fiber emission is $\sim 6.9\times10^7$~K, and the net fiber component (after the subtraction of background continuum) is on average 30--40\% of this, hence 2--$3\times10^7$~K. However, angular scattering by inhomogeneous structures in the corona may contribute significantly to the broadening of the apparent source size (which can amount to a few $\times 10''$), thus the intrinsic source size may be much smaller, or even point-like, which would results in a much higher brightness temperature and helps to maintain the coherency of the fiber emission.

We next obtained 2-D trajectories of fiber burst centroids by dividing them into different groups based on their drift rate distribution. These fiber trajectories seem to anchor near the footpoint of the hot post-flare loops (c.f., Figure \ref{fig3}). However, there appears to be no coherent loop-like structures shown in the AIA images along the exact locations of the derived fiber trajectories (c.f., Figure \ref{fig7}). There are two possible causes: (1) Similar to the argument proposed by \citet{2013ApJ...763L..21C} to account for type III radio burst trajectories invisible in EUV, although the loops in which fibers propagate are quite dense ($n_\mathrm{e}\approx 3\times 10^{10}$ cm$^{-3}$ as inferred from the observed radio frequencies), they may be too thin (100 km or less) to generate enough emission measure to be seen in EUV. (2) As the observed fiber bursts are probably due to fundamental plasma radiation, which suffers more from refraction due to density variation along the line of sight, the observed trajectories may be slightly shifted from the original source location, which results in a mismatch.

We further reconstructed 3-D fiber trajectories based on a barometric density model, whose parameters can be tweaked to optimize the fitting error between the observation-based velocities $V_{\mathrm{3D}}$ and the model-predicted velocities $V_{\mathrm{model}}$. We compared the fitting results using published models based on whistler waves, Alfv\'en wave solitons, and sausage-mode magnetoacoustic waves as plausible candidates for fiber burst exciter. Our investigation of fast- and slow-drift fiber groups at 18:40 UT suggests that, no matter which model is assumed, fibers in the dynamic spectrum with faster drift rate require not only a steeper density gradient (smaller density scale height) along the exciter trajectory, but also a faster exciter velocity itself. In addition, these two groups fail to be fit well using a single density model. The deviation in their trajectories also suggests that the fast and slow exciters may propagate along two different loops.

For the whistler wave model, a good fit can be achieved to with less than $10\%$ error in the three analyzed fiber groups. In the 1--2 GHz frequency range, the whistler wave packet can start at some coronal height around $\sim$~10 Mm. The corresponding group speed $V_\mathrm{g}$ is about 12,000 km$\cdot s^{-1}$. As it propagates upward in the corona, the exciter slows down to $\sim$ 4,000 km$\cdot s^{-1}$ at $\sim$36 Mm. The local magnetic field strength typically decreases from 62 to 8 Gauss, in which the Alfv\'en Mach number of the whistler speed $M_\mathrm{A}^\mathrm{w}$ is typically in the range $17\lesssim v_\mathrm{g}\left(\nu\right)/v_\mathrm{A}\left(\nu\right) \lesssim 28$.

For the MHD wave model, a small Mach number (typically $M_\mathrm{A}\leqslant 2$) renders an unphysically low source height (0--0.5~Mm above the photosphere), with marginal fitting errors at frequency above $\sim 1.8$~GHz due to the low predicted MHD velocity. However, the fitting becomes better with $M_\mathrm{A} = 3$ and also with a reasonable source height. Increasing the Mach number to an extreme value $M_\mathrm{A} =10$ only slightly improves the fitting by 1--2\%.

In conclusion, the fitting results from both models tend to favor a fiber burst exciter propagating at the speed comparable to the velocity of whistler waves. Of course, more observations along with the development of theories are needed to confirm the exact mechanism of fiber bursts.

\acknowledgments
\centerline{ACKNOWLEDGMENTS}
\hfill \break
We thank the anonymous referee for his/her constructive comments. We acknowledge support from NASA grants NNX14AK66G and NNH16ZDA001N and NSF Grants AST-1615807 and AGS-1654382 to the New Jersey Institute of Technology. We thank Dr. M. L. De Rosa for discussions on the PFSS extrapolation. The authors are grateful to VLA, RHESSI and SDO teams for providing the data used in this publication.

%% \allauthors
\listofchanges

\end{document}